\documentclass[vecphys]{svmult}

\usepackage{psfig}
\usepackage{epsf}

\usepackage{makeidx}         
\usepackage{graphicx}        
\usepackage{multicol}        
\usepackage[bottom]{footmisc}

\makeindex             

\begin{document}

\title*{Spectropolarimetry and the study of circumstellar disks}
\titlerunning{Spectropolarimetry} 
\author{Ren\'e D. Oudmaijer\inst{1}}
\institute{School of Physics and Astronomy, University of Leeds, LS2 9JT Leeds, U.K.
\texttt{roud@ast.leeds.ac.uk}}

%
%
\maketitle

\section*{Summary}

Circumstellar disks play an important role in many stages of the
evolution of stars. However, it is only possible to directly image
circumstellar disks for a few of the nearest stars. For massive stars,
the situation is even more difficult, as they are on average further
away than the more numerous low mass stars.  Here, we discuss
spectropolarimetry which is a technique that can reveal the presence
of disks on spatial scales of order stellar radii. It is therefore
extremely useful in studying disks around young stars, while the
method has prospects for quantitative modelling. In this review we
will discuss the status on the study of (accretion) disks around
intermediate- to high mass pre-main sequence stars and will
concentrate on the linear spectropolarimetry aspect. We wil make a
brief excursion to observations of massive evolved stars where
spectropolarimetry has revealed the clumpy structure of stellar winds,
and conclude with a brief outlook. \footnote{To be published as a chapter in {\it 'Diffuse Matter from Star Forming Regions to Active Galaxies' - A volume Honouring John Dyson}. Eds. T. W. Harquist, J. M. Pittard and S. A. E. G. Falle.  }

\section{Introduction}

The disk accretion paradigm has proven to be extremely successful in
explaining the formation of low mass stars. Due to angular momentum
considerations it is now commonly accepted that during gravitational
collapse a disk forms around a newly born star. The disk material
accretes onto the star by moving along magnetic field lines and
free-falling onto the stellar photosphere, giving rise to shocked
emission. This picture of magnetically controlled disk accretion has
been confirmed for the low mass T Tauri stars, where disks have been
found in high resolution images (Dutrey et al 1998) while evidence for
magnetic fields and magnetically controlled accretion via such disks
has been often found (Bertout 1989, Johns-Krull et al. 1999). This has
been reinforced by model simulations of the data (e.g. Muzerolle et al
2001).

For the ``intermediate mass'' (ranging from 2-8 solar masses) objects,
the optically visible Herbig Ae/Be stars, the situation is less
clearcut. Although a small number of Herbig Ae stars has now been
found to exhibit a (weak) magnetic field (e.g. Wade et al. 2005),
these more massive stars have radiative photospheres, and consequently
weaker, if any, magnetic fields. Therefore, the main component for the
T Tauri paradigm, magnetically controlled disk accretion, is less well
established and other scenarios that can form higher mass stars need to
be considered.  In addition, observationally, the presence of
accretion, and disk accretion in particular is much less settled for
these objects than for their low mass counterparts.  Spectroscopic
evidence for at least some transient accretion has now been found
(e.g. Mora et al 2004), mostly towards the A-type objects. Evidence
for the presence of extended, flattened disks has emerged recently
(e.g. Mannings \& Sargent 2000 in mm-CO observations, Fukagawa et al
2003, Grady et al 2001 in coronagraphic images). These large $\sim$
arcsec scale structures do not necessarily indicate the presence of
accretion disks, which, inevitably, are to be found at much smaller
radii.

The picture is even more obscure for the most massive stars, which we
loosely define as having masses greater than about 8-10 M$_{\odot}$.
Even a theoretical consensus has not emerged for these objects.  The
radiation pressure from a newly born, yet still accreting star, may
blow away the infalling material, halting further growth (e.g. Yorke
\& Kruegel 1977; Adams 1993). The maximum possible mass that can be
accumulated in simple accretion models is as low as 10M$_{\odot}$
(Wolfire \& Cassinelli 1987).  Disks around massive stars may help the
continued growth of these cores; Material accreting via a disk
captures less light than in the spherical case, while disks can
withstand the radiation pressure much better due to their higher
density (Norberg \& Maeder 2000; Yorke \& Sonnhalter 2002). Accretion
disks may thus be able to supply the necessary material to allow stars
to reach higher masses.  As an alternative, stellar coalescence due to
collisions of lower mass objects has been proposed as a way to
form a massive star (Bonnell, Bate \& Zinnecker 1998; Bally \&
Zinnecker 2005).

Next to the theoretical difficulties, observations of massive young
stellar objects (MYSO) are challenging.  The steepness of the Initial
Mass Function combined with their short lifetimes make them
scarce. They are therefore found at distances that are on average much
larger than those to lower mass stars, and this hampers studies aimed
at probing the circumstellar material close to the objects.  The
situation is made worse as most young massive stars are deeply
embedded in their natal clouds, in general preventing optical studies.
Thus, the presence of disks would go a long way into explaining the
formation of massive stars, but data on disks of any size are
sparse. Indirect evidence for the presence of, large scale, disks
comes from high resolution radio observations of the ionized
circumstellar material. Hoare et al (1994, 2002) have seen evidence
both for flattened disk-like winds and highly collimated jets,
suggesting the presence of inner disks feeding both processes. Among
the highest resolution studies of MYSOs we find the exceptional data
for the 8-10M$_{\odot}$ object G192.16-3.82. Shepherd et al. (2001)
obtained 30-40 milli-arcsec resolution data at 7 mm with the VLA for
their single target. Based on the similarities of the disk found at
100 AU with those around lower mass objects, they assume the disk to
be an accretion disk. Only recently have examples at similar scales
been published. Using the sub-millimetre array, Patel et al. (2005)
observed Cepheus A HW2, a 15M$_{\odot}$ pre-main sequence star and
found a disk or a flattened geometry with a size of order 330 AU.
Jiang et al. (2005) observed the Becklin-Neugebauer object, which has
an estimated mass of 7M$_{\odot}$.  They revealed a disk of a few
hundred AU with their adaptive optics assisted polarization images.
These studies are representative of the problems faced in the study of
the formation of young massive stars. Firstly, the
state-of-the-art-studies are single-object studies and secondly,
although disks with sizes comparable to the Solar system are observed,
it is extremely difficult to convincingly demonstrate the presence of
accretion. The latter can only be studied when we can trace structures
much closer to the star.

One such technique that is capable of probing the disks and immediate
circumstellar environment on very small scales, of order stellar
radii, is spectropolarimetry.  Indeed, the only evidence for disks
around intermediate mass pre-main sequence stars comes from
spectropolarimetry across the optical H$\alpha$ emission (Oudmaijer \&
Drew 1999, Vink et al 2002, 2005a).  In the case of HD 87643,
H$\alpha$ could be kinematically resolved and by comparison with basic
models (Wood et al. 1993) it was possible to show that the scattering
electrons were located in an expanding rotating disk (Oudmaijer et al
1998). But before discussing such results, let us first consider the
basics of spectropolarimetry below.

\section{Spectropolarimetry as a technique}

\begin{figure}[t]
\centering
\includegraphics[height=8cm,angle=0]{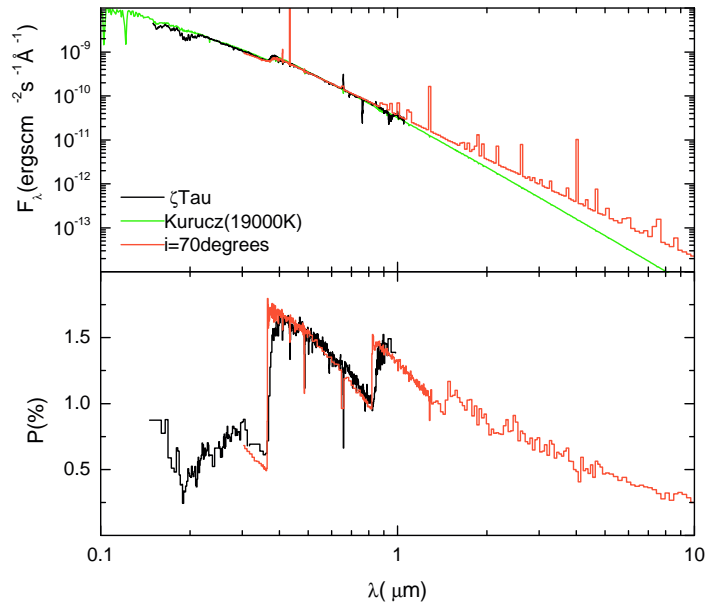}
\caption{The broadband polarization due to electron scattering. Data
of the Be star  $\zeta$ Tauri are shown. The upper panel shows the
spectral energy distribution and the lower panel the polarization. The
dotted lines represent the observations, while the solid line (up to
$\sim 1 \mu$m) is a model fit to the data.  Figure kindly provided by
J. Bjorkman.
\label{bjorkmod}}
\end{figure}

The principle of the method is rather simple: it uses the fact that
free electrons in an extended ionized region scatter the continuum
radiation from the central star and polarize it. If the projected
distribution of the electrons on the sky is circular, for example when
a disk is seen face-on, or when the material is distributed
spherically symmetrically, all polarization vectors cancel out and no
polarization is observed.  If the geometry is not circular as in the
case of an inclined disk for example, a net polarization of the
stellar continuum will be observed.  The polarization due to electron
scattering originates from the region closest the star where the
electron densities, and thus optical depths, are the largest. It is
typically sensitive to scales of a few stellar radii and results in
polarizations of order 1-2\% (Cassinelli et al. 1987). By measuring
the polarization due to ionized material, the method provides a means
to detect disks that otherwise can not be found. An example of the
broadband polarization behaviour around a star that is known to have
an inclined disk is shown in Fig.~\ref{bjorkmod}. The upper panel
shows the flux emanating from this system. The near-infrared excess
due to bound-free and free-free emission can be clearly seen. The
bottom panel shows the low resolution spectropolarimetry. The model
fits (solid lines - computed up to $\sim$1$\mu$m) reproduce the data
very well. As electron scattering is grey, it is not intuitive that
the observed polarization is not constant, but closely follows the
Balmer and Paschen jumps instead. This is due to the increased
hydrogen opacity shortward of the respective jumps, here the
electron-scattered photons will mostly be absorbed, reducing the
observed polarization, whereas the lower hydrogen opacity on the red
end allows more electron-scattered photons to escape (see e.g. Wood,
Bjorkman \& Bjorkman 1997). The magnitude of the jump depends on many
parameters, as for example the inclination of the disk.

Observing polarization on its own towards an object does not
necessarily imply the presence of an aspherical small scale structure
however. Aligned and elongated interstellar dust grains can
selectively absorb background light resulting in a net polarization
(Serkowski, 1962), while circumstellar dust in non-circular geometries
at larger distances from the star can also produce net polarization,
in much the same way as the free scattering electrons do.  As the
observed polarization can be a combination of electron-scattering,
circumstellar dust scattering and interstellar polarization, the
nature of the polarization can be hard to disentangle.  However, the
latter mechanisms have a different wavelength dependence of the
polarization than the fairly flat  polarization due to electron
scattering, and very broad wavelength coverage observations may help
in assessing the polarizing agent (e.g. Quirrenbach et al. 1997).
Yet, the ISP and circumstellar dust scattering can contribute the
majority to the observed polarization of an object. Unless the star is
closeby and does not have much circumstellar dust, as is the case for
the bright Be stars, the broad band behaviour is less conclusive than
we might hope for.

Here is where line spectropolarimetry comes in. It exploits the fact
that the hydrogen recombination line emission arising from within the
ionized material will be polarized much less efficiently than the
photospheric continuum.  This is because of two reasons. Firstly the
line emission will encounter less electrons as it travels a shorter
distance through the disk. Secondly the lines can form over a larger
volume than the electron scattering region, which is very much
confined to the inner, denser, regions. This implies that emission
lines will be less polarized than the continuum or even unpolarized
altogether. Such a ``line-effect'' will be absent for the interstellar
polarization as dust grains have a broad wavelength dependence, while
circumstellar dust typically resides at much larger distances than the
hydrogen recombination line forming region and consequently both star
light and line emission will be scattered equally.

Therefore, when interpreting polarization spectra, a change in the
polarization spectrum across an emission line such as H$\alpha$
immediately indicates the presence of a flattened structure close to
the star. This technique can thus reveal the presence of aspherical
structures at scales that can not even be imaged with the most
sensitive telescopes. Indeed, it is distance independent, so
potentially stars in other galaxies can be studied in much the same way
as closeby Be stars. 


\subsection{An example : the Be star  $\zeta$ Tau}

\begin{figure}[t]
\centering
\includegraphics[height=8.5cm,angle=-90]{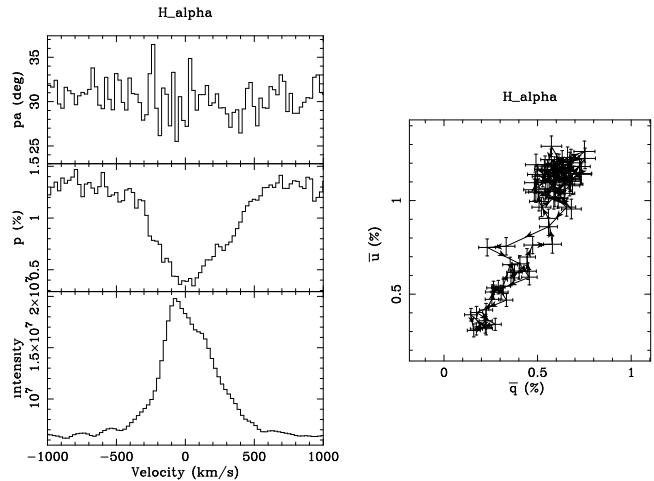}
\caption{H$\alpha$ spectropolarimetry of $\zeta$ Tau.  On the left
 hand side we show the so-called ``triplot''. This presents the
 intensity spectrum (bottom), and the polarization percentage (middle)
 and polarization angle (top) rebinned to a given error. The right
 hand graphs display the polarization in {\it QU} space at the same
 binning.  Image credit : Mottram et al. 2007}
\label{zetalfa}       
\end{figure}

Spectropolarimetry was already explored in the seventies by Poeckert
\& Marlborough (1976). They used narrow-band polarimetry to prove that
Be stars are surrounded by ionized disks. It was only many years later
that this result could be directly confirmed with high spatial
resolution radio and optical interferometry (Dougherty \& Taylor 1992,
Quirrenbach et al. 1994, respectively).  With the advent of CCD
detectors and the installation of stable and efficient polarization
optics, it has become possible to make routine spectropolarimetric
observations at medium to high resolution.

Data of the H$\alpha$ line of a typical Be star, $\zeta$ Tau, taken in
January 1996 is presented in Fig.~\ref{zetalfa}, where the Stokes I
(direct intensity) vector is plotted in the bottom panel, and the
polarisation percentage and polarisation angle (PA) are displayed in
the middle and upper panels respectively. These data are adaptively
binned such that individual bins correspond to 0.1\% in
polarization. The line is asymmetric, assuming the line to be doubly
peaked, the blue component stronger than the red component.  The
continuum polarisation of 1.3\% and the PA of $\sim$30$^{\rm o}$
follow the trend of increasing polarization observed over the past
years as noted by McDavid (1999). The polarisation across the
H$\alpha$ line shows a marked drop with respect to the continuum and
is a clear example of the ``line-effect'' revealing the presence of
aspherical symmetries, i.e. the disk.

The right hand panel shows the {\it QU} polarization vectors plotted
against each other, and the line excursion is clearly present as
well. The cluster of points in the upper right hand corner of the
graph represent the continuum polarization ($p^2 = Q^2 + U^2$) and the
excursion to the bottom left shows the polarization when the
wavelength moves to and from the line center.  The amplitude of the
excursion, about 1\%, is a measure of the polarization due to
electron scattering and in good agreement with theoretical
expectations. The individual datapoints are distributed around the
linear excursion, and its pattern is not well understood. However, the
scatter appears to be of order the 0.1\% error binning applied and
care should be taken when trying to interpret such changes.

The data also allow us to directly determine the {\it intrinsic}
polarization angle (PA) from the excursion across the line profile
observed in the {\it QU} diagram. The PA can be written as $\Theta =
\frac{1}{2}\times$ atan($\Delta$U/$\Delta$Q). This results in 32$^{\rm
o}$ with an estimated experimental error of around 4$^{\rm o}$.  In
the case of optically thin scattering, the situation encountered the
most, the polarization angle will be perpendicular to the disk
structure itself, and the intrinsic polarization angle implies a disk
orientation of -58$\pm 4^{\rm o}$ on the sky.

\begin{figure}[t]
\centering
\includegraphics[height=7.5cm,angle=0]{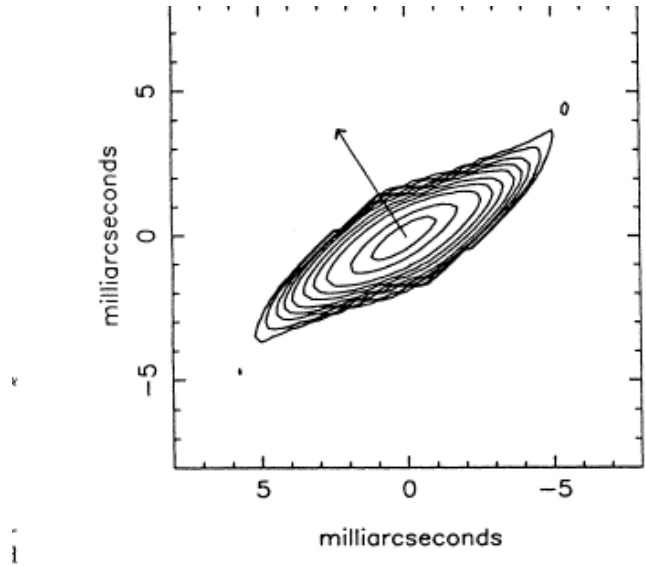}
\caption{Quirrenbach et al.'s (1994) maximum entropy reconstruction of
$\zeta$ Tau in the H$\alpha$ emission line. The arrow indicates the
position angle of the linear polarization. Credit : A\&A}
\label{zetaq}       
\end{figure}

For comparison, we show Quirrenbach et al.'s (1994) high resolution,
reconstructed, image in the H$\alpha$ line of $\zeta$ Tau in
Fig.~\ref{zetaq}.  With a {\it V} band magnitude of 3 and a distance
of 128 pc (as derived from the Hipparcos parallax of 7.8 milli-arcsec)
this is one of the brightest and nearest objects in the sky. Yet, the
extent of the disk is only a few milli-arcseconds. The second contour,
counted from the centre, traces 50\% of the peak light, and implies a
full-width-at-half-maximum (FWHM) of 5 milli-arcsec. Quirrenbach et
al. (1994, 1997) measure the disk's position angle to $-58\pm4$$^{\rm
o}$.  This value is in very good agreement with the position angle
derived from the spectropolarimetry and presents a compelling
validation of spectropolarimetry as a means to reveal small scale
asphericities.

\begin{figure}
\centering
\includegraphics[height=15cm]{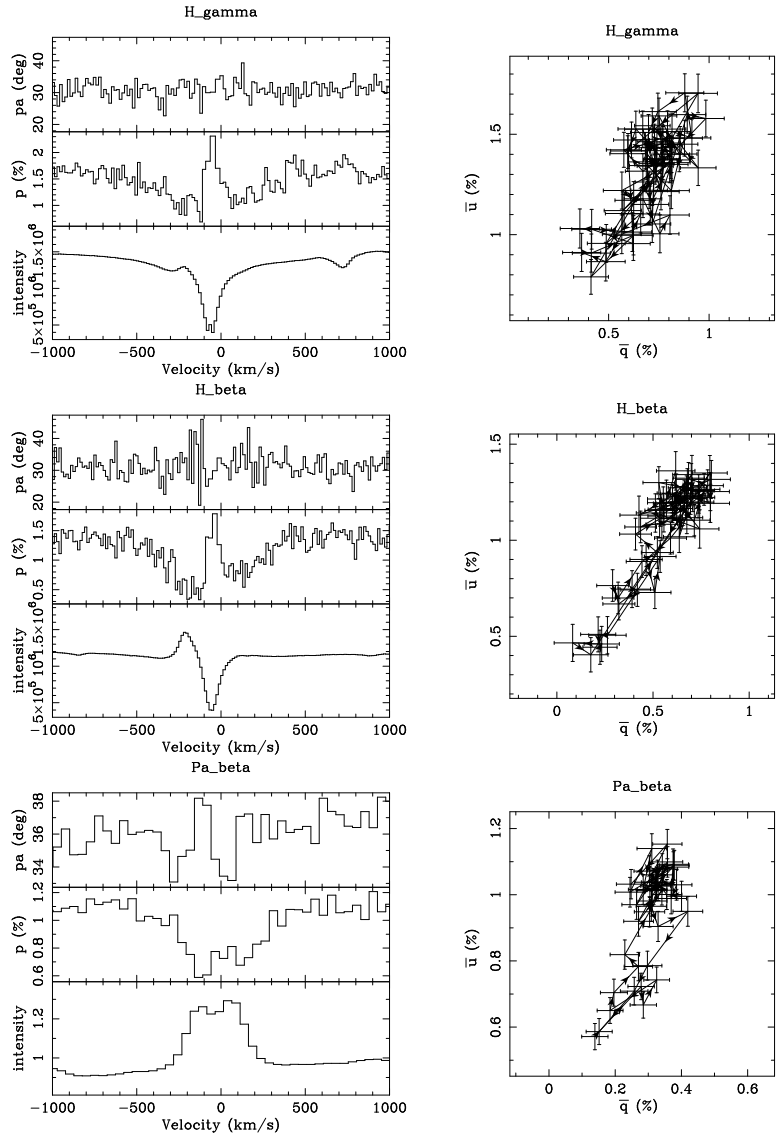}
\caption{Medium resolution spectropolarimetry of the Be star $\zeta$
Tau. For each of the hydrogen recombination lines, H$\gamma$,
H$\beta$, and Pa$\beta$, two figures are shown as in the previous
figure.  Never before have data such as this, covering this many lines
ranging from 4340$\rm \AA$ to the near-infrared (1.28 $\mu$m) been
published as no comparable data set is in existence.  Credit for the
optical data : Mottram et al. (2007), Near-infrared data : Oudmaijer
et al. (2005).}
\label{zeta}       
\end{figure}

Let us now extend the wavelength coverage by a factor of a few.  In
Fig.~\ref{zeta} three other hydrogen recombination lines in the
spectrum of $\zeta$ Tau are shown, from top to bottom ordered in
wavelength, the optical H$\gamma$, H$\beta$ lines and the
near-infrared 1.28$\mu$m Pa$\beta$ line.  The optical lines were
observed 8 years after the H$\alpha$ line shown in
Fig.~\ref{zetalfa} (September 2004, Mottram et al.
2007). The Pa$\beta$ line was obtained in September
1999 (taken from Oudmaijer, Drew and Vink 2005).  As  expected
from their lower transition probabilities the lines are much weaker
than H$\alpha$.  The optical lines show stronger blue peaks, like
H$\alpha$. Pa$\beta$ was taken at a different epoch and shows both
peaks at similar strength. Kaye \& Gies (1997) observed the same for
H$\alpha$ and clearly, the profiles are variable. Such blue peak to
red peak variability of the line-profile has most often been
interpreted as being due to the presence of one-armed density waves
within the circumstellar disk (see e.g. Porter \& Rivinius 2003). Note
also that the optical, simultaneous, continuum polarization is fairly
flat, as expected from electron-scattering (see Fig.~\ref{bjorkmod})

The intrinsic angle derived from the {\it QU} graphs is 29$^{\rm o}$,
29$^{\rm o}$, and 32$^{\rm o}$ for the lines respectively. Both
H$\gamma$ and H$\beta$ have consistent values, but the H$\alpha$ and
Pa$\beta$ data, taken a few years earlier, are different by a few
degrees. This difference could be explained by the generous errorbars,
but may also be a confirmation of the one-sided arm hypothesis of the
disks. A axi-symmetric disk can display polarization changes, due for
example to changes in its density, but it will normally be expected to
show the same orientation. On the other hand, orbiting density
enhancements could induce variations in the orientation if they are
not always located in the same plane.

A special word on the weakest line in the graph.  The H$\gamma$
emission does not even exceed the continuum level, yet, a line-effect
is clearly visible in the data. In addition, the position angle
derived from the excursion in the {\it QU} plot is perfectly
consistent with the other, stronger, lines.  It is because we have
knowledge of the other lines in the spectrum of $\zeta$ Tau that we
can be confident that a true line-effect is present.  This, certainly
at first sight puzzling, fact that we can see this effect at all is
because the disk emission is still substantial. The photospheric
hydrogen lines of a B2IV star such as $\zeta$ Tau are strongly in
absorption. Any emission first has to fill in the strong absorption
line before exceeding the continuum.  This emission is unpolarized and
significant compared to the underlying remaining photosheric
radiation, and gives rise to a line-effect.  It is a testament to the
spectropolarimetric method that such a weak line still allows a disk
to be revealed. This is the first published instance of a
``line-effect'' in spectropolarimetry, without the presence of an
emission line!

In summary, line spectropolarimetry is a powerful method to reveal the
presence of circumstellar disks at very small scales.  In the
following we give a brief overview of the most recent developments in
the field of spectropolarimetry. Inevitably, the topics will be
slightly biased towards the interests of the author, and will focus
mainly on pre-main sequences star, a brief excursion to
clumpier material will be presented later.

\section{Application to Pre-Main Sequence Stars}

We now move towards the pre-main sequence stars. As discussed earlier,
one of the major issues in the field of star formation is whether
these objects accrete material via a disk or not. Establishing the
presence of small scale disks will substantially contribute to settling this
issue; to have disk accretion, we first need disks reaching into, or
close to, the stellar photosphere. On the longer term, follow-on
models have the potential obtain physical parameters for such disks.

Linear spectropolarimetry requires comparatively high spectral
resolution to properly sample the emission line profiles and very high
signal-to-noise ratios. The binning in the previous figures was done
to an accuracy of 0.1\% in polarization, which corresponds to
measuring the spectrum to a precision of one-thousandth of the total
collected light. This means that signal-to-noise ratios (SNR) of order
1000 are needed. If we then also add the requirement of moderately
high spectral resolution, it may be clear that we are restricted to
observing optically bright objects. In the 4m telescope era (AAT,
WHT), objects brighter than about {\it V} = 10 are routinely observed,
objects in the 11-12 magnitude range have been observed, but these
become very challenging.

Such bright limits prevent us from observing the most massive, heavily
embedded, optically obscured pre-main sequence stars. Having said
that, an excellent starting sample to address the issue are the Herbig
Ae/Be stars. These objects are intermediate between the lower mass T
Tauri stars and the high mass stars. They therefore provide a
continuous coverage of the mass spectrum and will allow us to mark the
switch from magnetically controlled accretion to other
mechanisms. Last but not least, the most massive objects among the
Herbig Be stars ($>$8-10M$_{\odot}$) are already direct examples of
objects which should not have formed via spherical accretion alone.


\subsection{On the presence of the effect}

\begin{figure}[t]
\centering
\includegraphics[height=7.15cm]{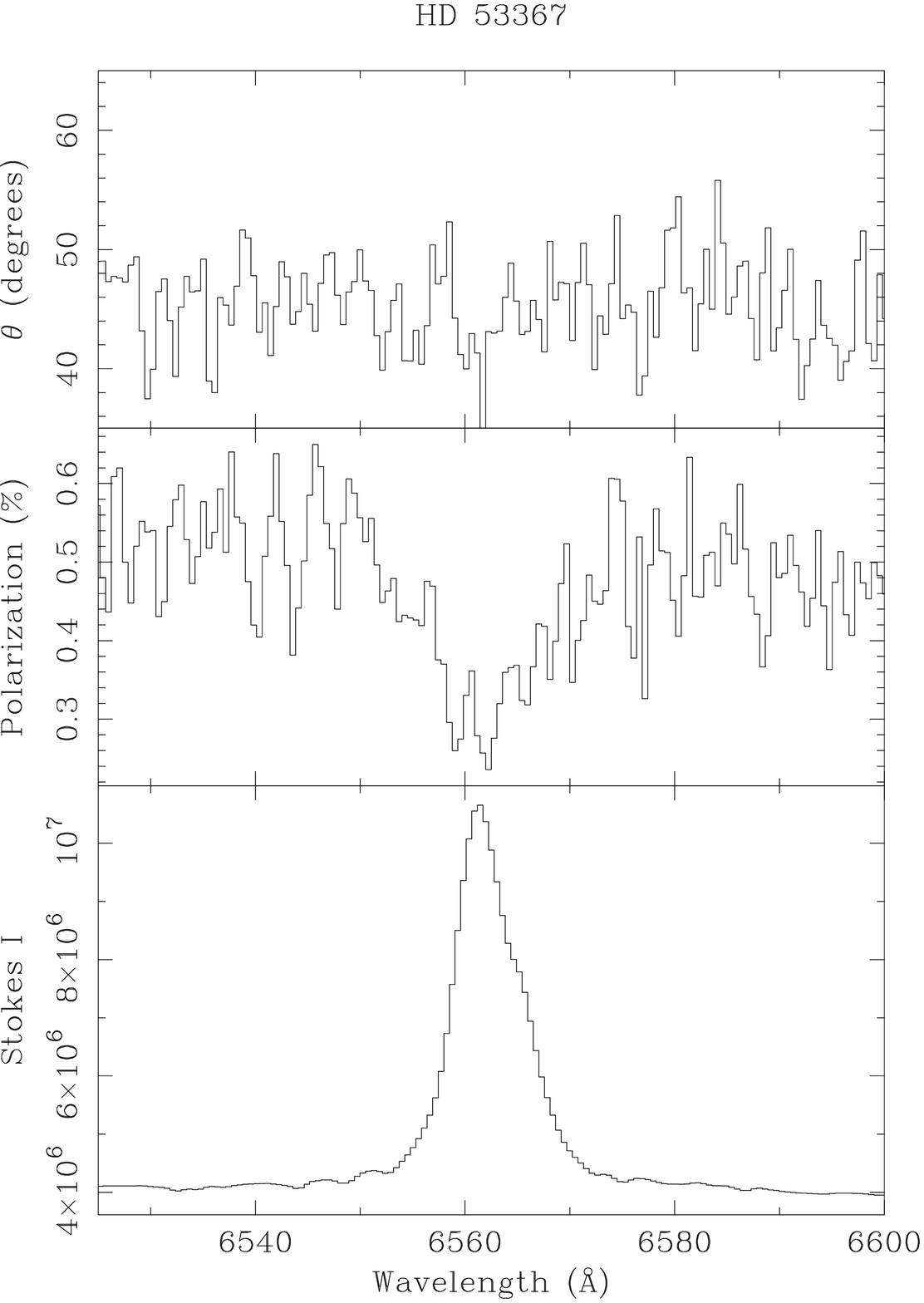}
\includegraphics[height=7.15cm]{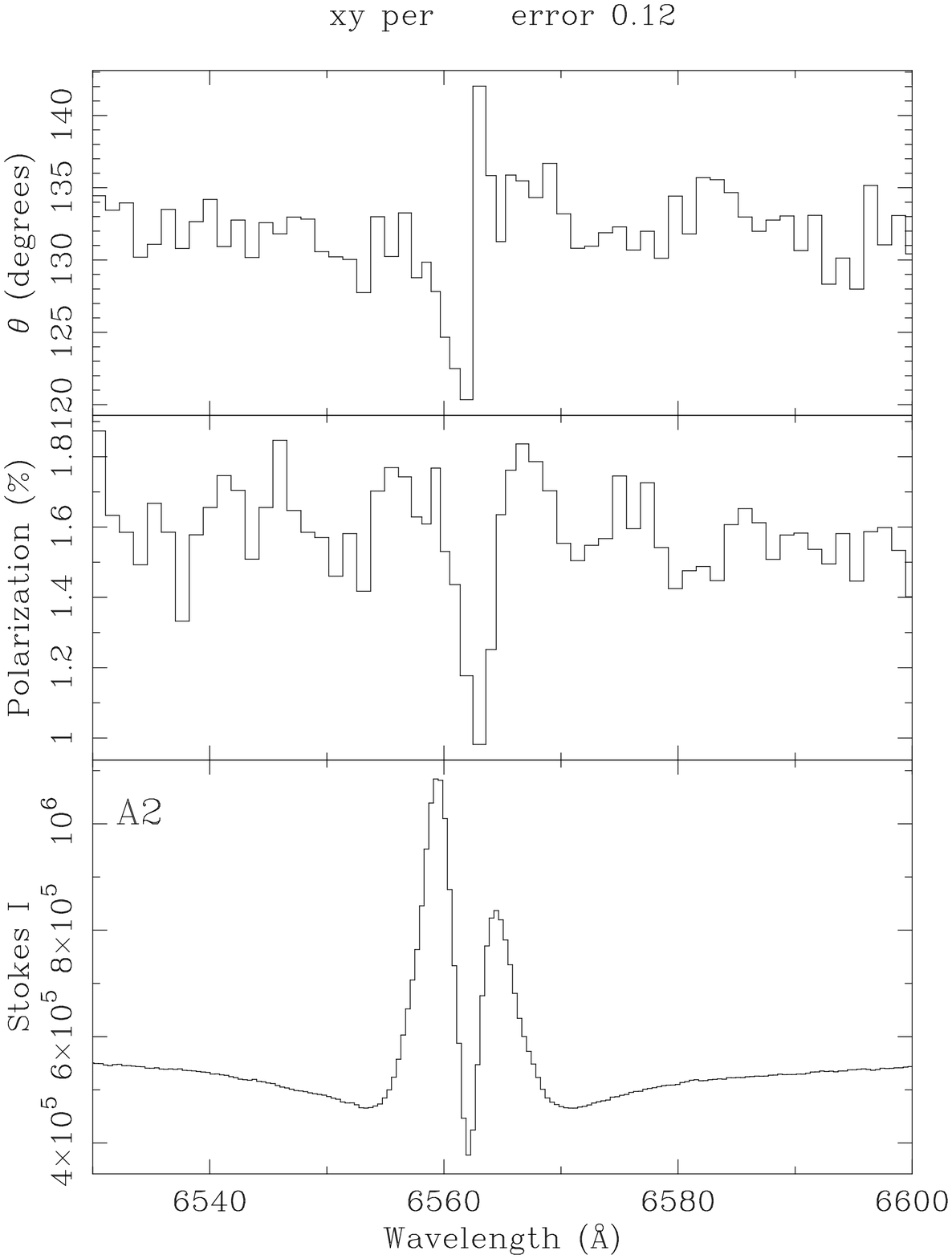}
\includegraphics[height=7.15cm]{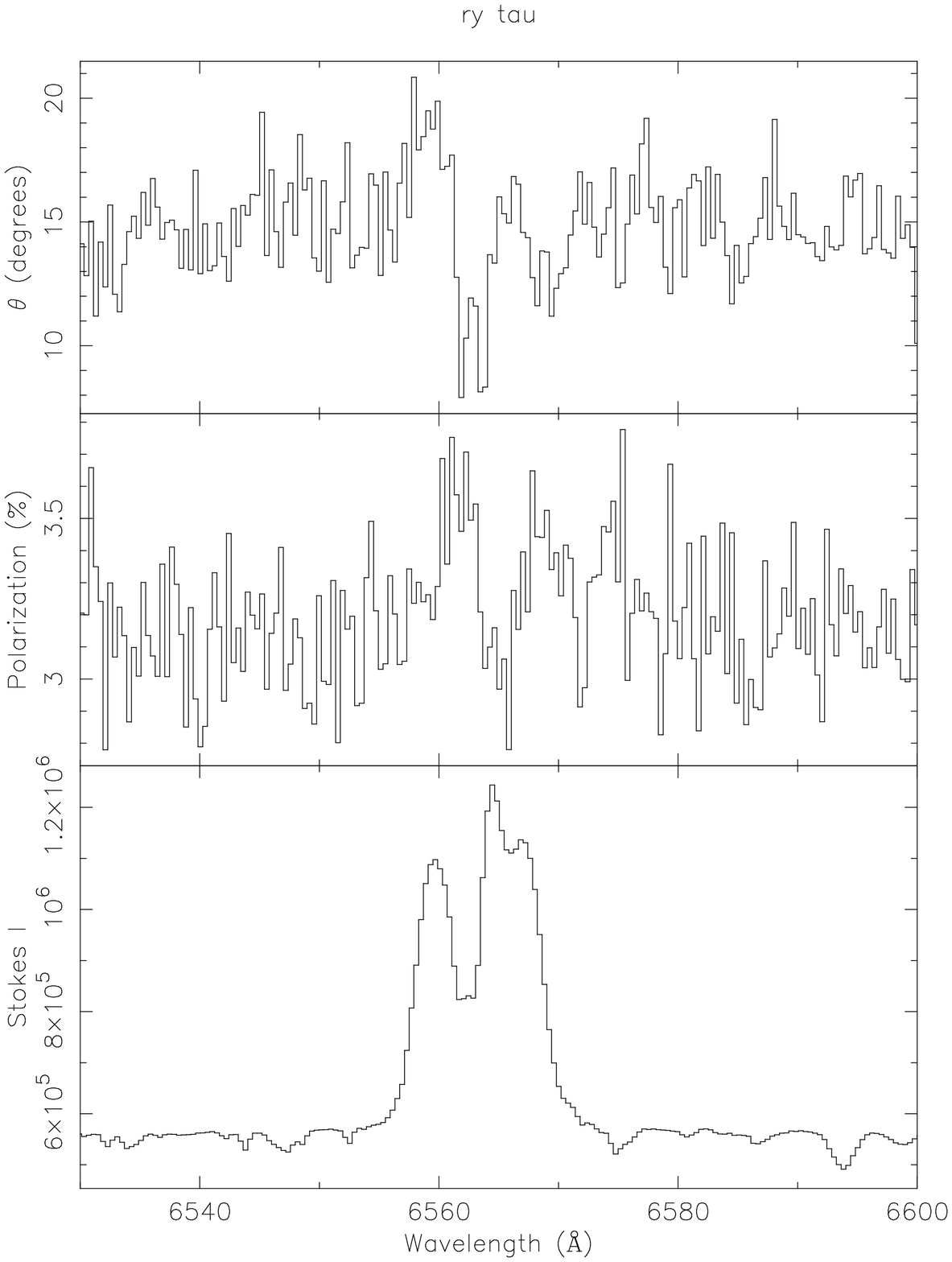}
\caption{Spectropolarimetry of a Herbig Be object (top left), a Herbig Ae
star (top right) and a T Tauri star (bottom). As discussed in the text,
there is a large resemblance in the spectropolarimetric properties
between the Herbig Ae and T Tauri stars.}
\label{three}       
\end{figure}

Although line spectropolarimetry had been performed for a selected
number of evolved stars such as AG Car (Schulte-Ladbeck et al. 1994
and references therein), in the early nineties, data on young stars
was sparse. The best data are arguably those of Schulte-Ladbeck et
al. (1992) of the Herbig Ae/Be star HD 45677. Their emphasis was on
the broadband behaviour of the spectropolarimetric data however, and
the spectral resolution was not sufficient to detect a line-effect in
the H$\alpha$ emission line.  The first medium resolution ($\sim100$
kms$^{-1}$) data were presented by Oudmaijer \& Drew (1999) and Vink
et al. (2002) who observed a significant sub-sample from the Th\'e et
al. (1994) catalogue of Herbig Ae/Be objects. Some examples of
line-effects are shown in Fig.~\ref{three}, where also a T Tauri star
from Vink et al. (2005a) is plotted for comparison.  Whenever data on
larger scale disks are available, often obtained after the
spectropolarimetry, the intrinsic polarization angles are perpendicular to the
position angles of the imaged disks, as expected.

To within the sensitivity, more than half of the two dozen objects
surveyed show a line effect (16 out of 23 objects).  As the systems
are oriented randomly in the line of sight it is inevitable that some
objects are face-on or close to face-on.  A non-detection does
therefore not necessarily imply the absence of a disk, because a
face-on disk would appear circular on the sky, and not produce a
line-effect. In the slightly inclined case, only very small effects
are present and these may be hard to pick up at the 0.1\% sensitivity
level.  Taking this into account, the high detection rate of a
line-effect (70$\pm$17\%) strongly suggests that all systems are
surrounded by disks.  This survey therefore provides evidence that the
disk accretion scenario is a strong contender to explain the formation
of massive stars.

It could be argued that the mere presence of a disk alone can not be
regarded as the smoking gun for disk accretion scenarios. This
objection is equally valid for the disks noted in the
spectropolarimetry as it is for the high resolution imaging studies by
e.g. Shepherd et al. (2001).  Recent studies in binary formation may
help in this respect.  Studies of multiplicity amongst Herbig Ae/Be
stars have retrieved large binary fractions up to 70\% (Baines et
al. 2006, Leinert et al. 1997).  As reviewed by e.g. Clarke (2001),
the formation of a binary system is due to the break-up of the
pre-natal cloud into two or more fragments, or due to the capture of a
lower mass star.  These two competing models predict a different
alignment of the binary systems and the disks around the
stars. Capture models result in randomly oriented disks with respect
to the binary position angle (e.g. Bally \& Zinnecker 2005), while
fragmentation models predict co-planar disks around the stars (see
also Wolf et al. 2001 who find this for T Tauri stars).  Baines et
al. (2006) compare the orientations of the binary Herbig Ae/Be at
their disposal with the intrinsic polarization angle if present in the
literature. Five of the six objects that have data for both, have
intrinsic position angles perpendicular to within 25$^{\circ}$ from
the binary position angle.  From a statistical analysis they reject
that the sample is drawn from a population of randomly aligned
disk-binary systems at the 98.2\% level. Instead, as the intrinsic
polarization angle is perpendicular to the disk orientation, this
result indicates that the circumprimary disks and the much larger
binaries are well aligned.

This provides strong evidence in favour of the fragmentation scenario
for the formation of binary systems as it predicts aligned disks, and
argues against the stellar capture scenario or the stellar merger
theory for the more massive stars.

\subsection{Resolved emission lines}

As discussed above, the classical idea of a depolarization across the
emission line is due to the electrons scattering the stellar
photospheric emission, and not the line itself.

However, for a large fraction of the stars, more than just a simple
depolarization signature is seen, and even instances of enhanced
polarization are observed.  This occurs more often in the cooler
Herbig Ae stars than in the Herbig Be stars.  Intrigued by this, Vink
et al. (2003, 2005a) observed a sample of T Tauri stars to investigate
whether the line polarization properties depend on stellar mass, and
if so, at what spectral type there may be a change in properties,
perhaps sign posting a different formation mechanism.
Fig.~\ref{three} shows a representative sequence with data from Herbig
Be star, a Herbig Ae star and a T Tauri star. The earliest type star
shows a clear depolarization. The T Tauri star has an additional
polarization across the line, and despite many attempts to correct for
intervening dust polarization, the enhanced polarization across the
line remains. The same applies to the Herbig Ae object.  Vink et
al. (2005a) quantify these different line profiles using a measure of
the width of the polarization feature across the line.  There appears
to be a trend in that the ``shape'' of the polarization across the
line in the Herbig Be stars is broader than that of the T Tauri stars,
with the Herbig Ae stars intermediate between the two, as also visible
in the Figure. The {\it QU} graphs also show the marked difference,
whereas the Herbig Be object displays a linear excursion over
H$\alpha$, the later type objects can best be described with a
``loop'' (Vink et al. 2005b).

The lack of a true line depolarization in T Tauri stars can be
explained by the fact that their disks are magnetically truncated and
have an inner hole. As most electron-scattering occurs in the inner
regions of the disk, the polarization of the photosphere of a T Tauri
star due to electron-scattering could be very low.  As a result, no
true line effect would be expected in in the spectropolarimetry of T
Tauri stars.  The enhanced polarization can be explained as being due
to a compact source of emission that scatters off a rotating disk-type
structure.  The best candidates for the compact sources will be the
accretion hot spots where the accretion flow free falls onto the
stellar surface.

The similarity in the spectropolarimetric behaviour of the convective
T Tauri stars and their hotter Herbig Ae counterparts led Vink et
al. (2003) to suggest that magnetically controlled accretion plays a
role in stars more massive than T Tauri stars. In fact, this was the
first indication that the change from magnetically controlled
accretion to other processes occurs at higher temperatures than
previously thought. Not much later, Hubrig et al (2004, 2006) and Wade
et al (2005) detected, for the first time, magnetic fields in some
Herbig Ae stars, supporting this idea.

\begin{figure}[t]
\centering
\includegraphics[height=7.15cm]{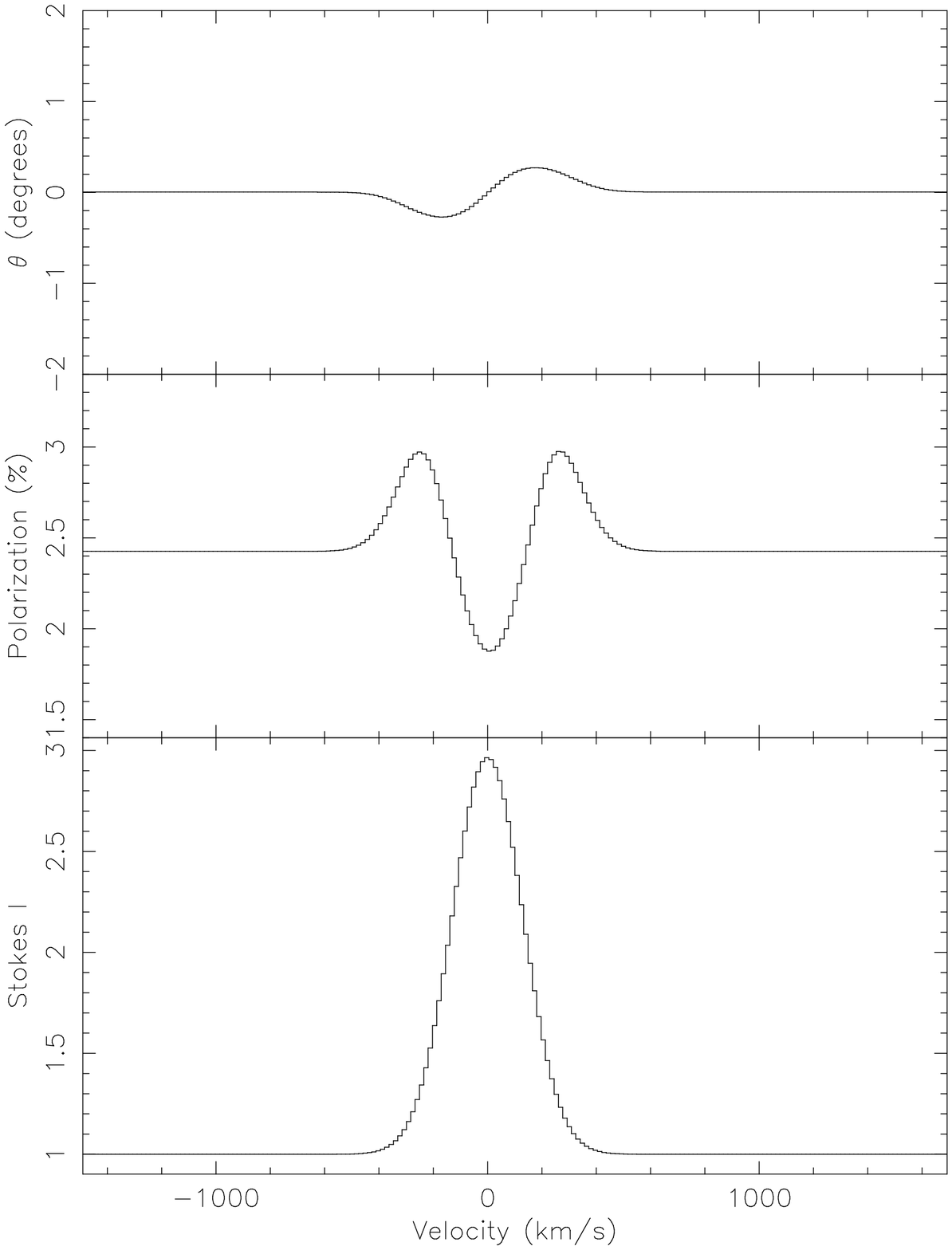}
\includegraphics[height=7.15cm]{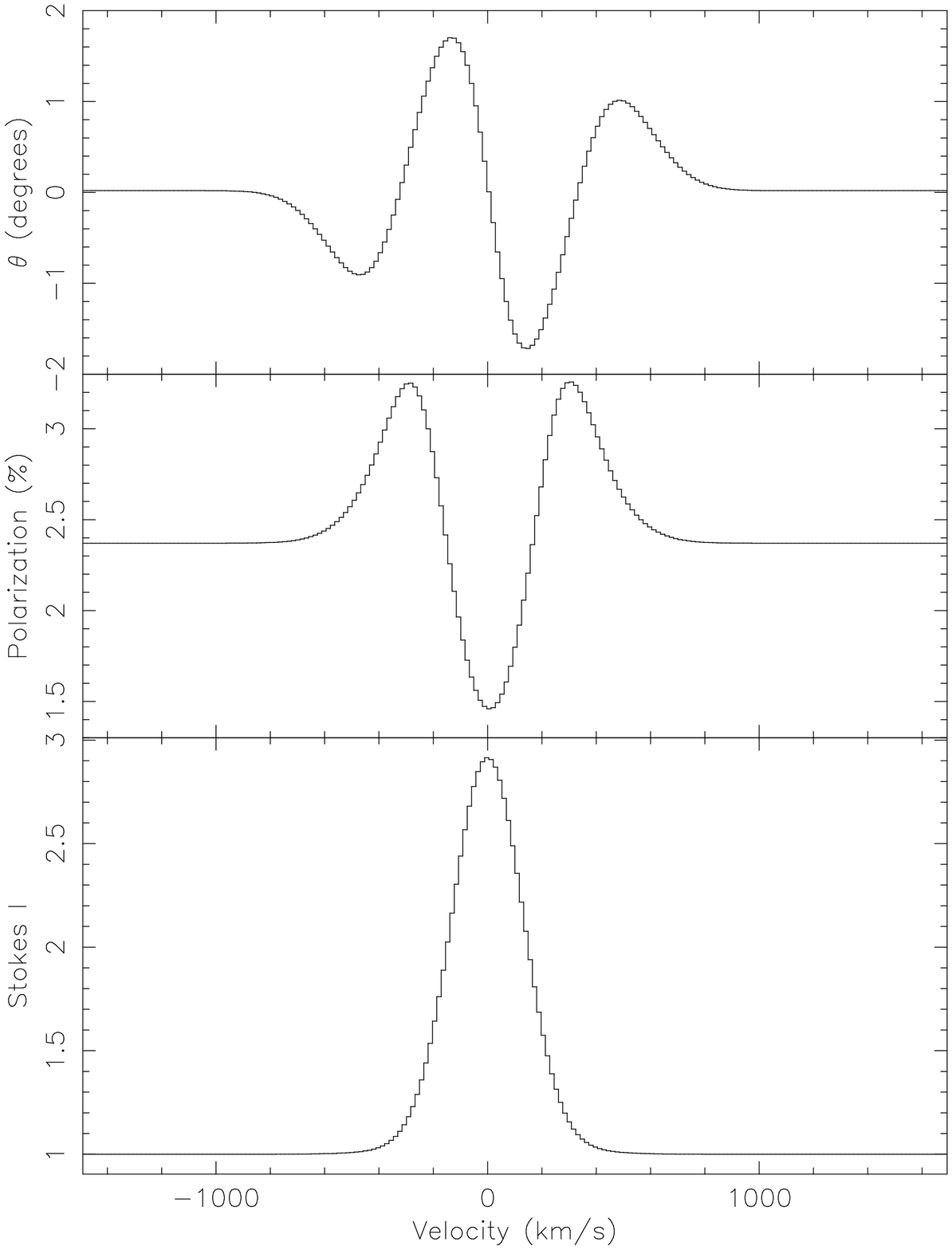}
\caption{Model computations of the line polarization in the case of a
finite star and a disk with an inner hole (left) and without an inner
hole (right). Note the double flip in rotation revealing that the
disks reaches into the stellar photosphere. Figure kindly provided by
Vink. }
\label{jormod}       
\end{figure}

\subsection{Modelling : towards physical parameters of the disks}

So far, we have only discussed the qualitative, observational, result
that disks are present around the target objects.  In parallel with
the improved instrumentation, theoretical models have matured as well.
The ultimate aim of such models is to constrain the astrophysical
parameters of the disks giving feedback to the models of the disks'
formation.  In order to reproduce the observed polarization, line
emission and the spectropolarimetric line profiles, one needs
full-blown radiative transfer models of such disks that take the
dynamics of the system into account as well.  At present, this is not
readily possible, and currently the problem is approached from several
directions.  Broadly speaking there are two families of models being
developed to interpret spectropolarimetric data.

The first line of attack is based on the line-profiles themselves. The
resulting polarization as a function of wavelength gives valuable
kinematical information on the velocity field of the scattering
electrons.  Vink et al. (2005b), using the Monte Carlo code of Harries
(2000) and building on the analytical work of Wood et al. (1993)
consider the scattering of radiation from a central source by rotating
disks. The line emission by being scattered off the electrons within
the disks will not only be polarized, but also carry the Doppler
information with them. By simple comparison with the Wood et
al. (1993) models, Oudmaijer et al. (1998) were able to infer from
their observed line profiles that the scattering material was located
in an expanding, rotating disk.  A remarkable new result from the Vink
et al. (2005b) paper is that it proves possible to check whether the
disk reaches into the (finite) star or whether an inner hole is
present, as expected for the magnetically truncated disks in T Tauri
stars. An example of their model results is shown in
Fig.~\ref{jormod}. The left hand panel shows the polarization profiles
for a disk with a hole, a rotation is present in the polarization
angle. The right hand panel displays the case for the disk reaching
into the star. The polarization itself has qualitatively the same
shape, but due to light being eclipsed by the stellar disk, there
occurs a double rotation in the polarization angle. This appears to be
a robust result, and will allow us to pinpoint the nature of the
disks, well before imaging data can resolve them. In fact, when
the inclination of the (larger scale) disk is known, it proves
possible to estimate the size of the inner hole. Vink et al. (2005a)
apply this new technique to a sample of T Tau stars and find that the
inner hole of the disk around SU Aur is larger than 3 stellar
radii. This is consistent with the interferometric findings of Akeson
et al. (2005) who fit ring and disk models to their visibility data of
the object, and for the ring model derive an inner radius of 0.18 AU.
    
This approach is particularly powerful in interpreting line profiles
from compact emission sources, as for example the accretion hot spots
from T Tauri stars and other general cases. The natural next step is
to incorporate radiative transfer in the models to simulate not only
the scattering but the line emission itself as well.

This method is perfectly complemented by the radiative transfer models
of circumstellar, dense, ionized disks. These models are becoming
increasingly increasingly realistic (e.g. Carciofi \& Bjorkman
2006). Such 3-D Monte Carlo models not only solve the radiative
transfer but by following the energy packets as they are created and
scattered, also compute the observed polarization (see
Fig.~\ref{bjorkmod}, cf. Wood et al. 1997, note that the line effect
is readily visible). Such models, applied to the broad wavelength
spectropolarimetry have already allowed the determination of the disks'
opening angles (2.5$^{\rm o}$ in the case of $\zeta$ Tau), quantities
which are impossible to measure by any other observational means.  The
natural next step is to implement the line profiles in the codes and
it will be only a matter of time that the exciting prospect of full
models of line spectropolarimetry is reality.

\section{Variability studies}

\subsection{UXOR variability}

An often overlooked property that reveals an object is intrinsically
polarized and therefore surrounded by aspherical material is
variability in the observed polarization. Because interstellar
polarization is not expected or observed to change on short
timescales, any variability immediately points at either circumstellar
dust or electrons being the polarization agents.

A sub-sample of pre-main sequence stars displays a special type of
photo-polarimetric variability, commonly referred to as the `UX Ori'
phenomenon (e.g. Grinin et al. 1994, Oudmaijer et al. 2001).  From
long term monitoring of a small number of young stars, Grinin et al.
(1994) identified a group of objects that are photo-polarimetrically
variable. This group of stars shows increased polarization when the
optical light of the stars is faint. Crucially, the objects also are
redder at fainter magnitudes, while in extreme visual minima there is
a colour reversal, the observed colours become bluer again.  Named
after their proto-type, UX Orionis, these stars are commonly referred
to as UXORs. UX Ori itself is  a well-known Herbig Ae/Be star, and
indeed many UXORs fall in the intermediate mass Herbig Ae/Be category.

The main explanation of this phenomenon concerns the existence of dust
clumps located in a disk-like configuration rotating around the star.
When the dusty clumps are not in our line of sight, the star will be
observed at maximum light, with only a slight contribution of
radiation scattered off the dusty disk.  If the dust intersects
the line of sight, light from the star will be absorbed, and the
relative contribution of the scattered light to the total light
increases, increasing the observed polarization. The fact
that the reddening of the star coincides with the faintening, leaves
little doubt that dust absorption indeed plays the main role in the
process. In cases of extremely deep minima, the light from the star is
blocked almost entirely, resulting in a `blueing' of the energy
distribution, as now mostly scattered light dominates the observed
light.  Depending on the distribution of dust-clouds, the light can be
more or less absorbed during a period of photo-polarimetric
monitoring. A direct observational consequence is that any variations
in {\it QU} space predominately occur along a straight line with a
slope perpendicular to the orientation of the disk, and is observed.

A more refined explanation of the UXOR behaviour is provided by
Dullemond et al. (2003). They consider that the dusty disks around
UXORs are thin and that the inner rims of the disks shadow the outer
parts. The inner rim of the disks is heated by the stellar radiation
field and is ``puffed up''.  Dullemond et al. discuss and model the
case where only slight hydrodynamical perturbations can result in an
uneven inner rim, and such fluctuations in the density can
occasionally obscure the stellar light from view, in exactly the same
manner as the qualitative clumps discussed previously in the
literature.

\begin{figure}[t]
\centering
\includegraphics[height=6cm,angle=0]{./agcar_tri2.ps}
\includegraphics[height=6cm,angle=0]{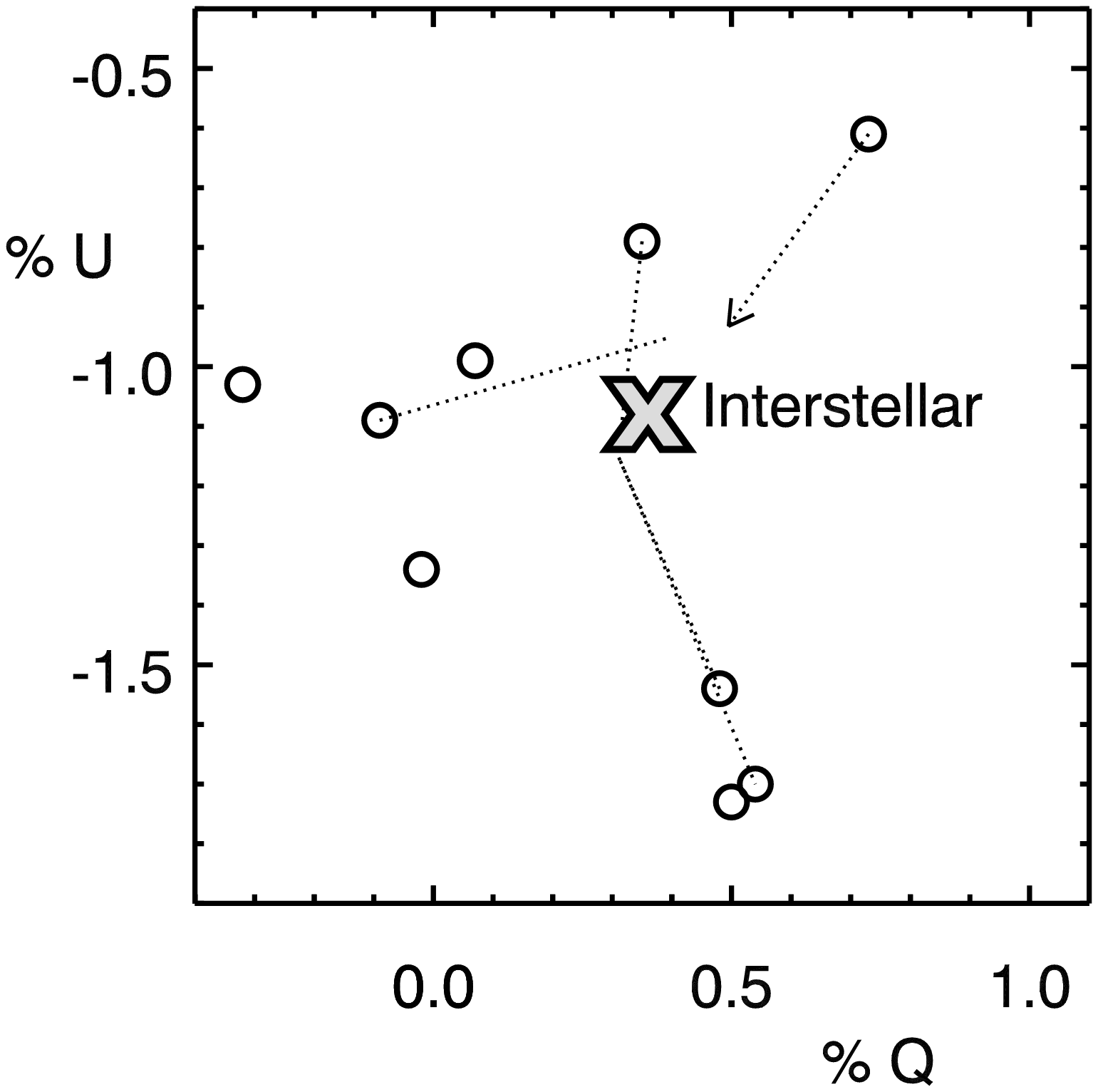}
\caption{
The polarization spectrum across H$\alpha$ of the Luminous Blue
Variable AG Car. The data are represented as in the previous
figures. The right most panel shows the {\it QU} graph where existing
data are plotted. The big cross in the center denotes the polarization
measured in the line centre, which is fairly constant over time, while
the other points are the continuum polarization measurements. It is
clear that the objects is variable. Figure kindly provided by Ben
Davies.
\label{lbv}}
\end{figure}

\subsection{Going clumpy : Luminous Blue Variables}

Adding complications to simple observational pictures has always been
a habit of theoreticians. Introducing clumps in stellar winds
previously assumed to be smooth is but an example.  I first met John
Dyson at a workshop on evolved stars in Chile in 1991. In those days,
the progenitors to Planetary Nebulae, post-Asymptotic Giant Branch
stars, were not very numerous, mostly single and surrounded by
spherically symmetric material, or so we thought. At that time,
annoying complications such a circumstellar disks and binary central
stars were creeping up already (e.g. Waters 1991), but John Dyson -
who gave one of the most hilarious after dinner speeches I've ever
witnessed - managed to complicate the issue even further with a
flamboyant presentation on clumpy winds (Dyson 1991). I actually
proceeded approaching the problem by considering disks around evolved
stars. A step up in complexity from the spherically case at least.
Fast forward more than a decade, and we arrive at the more massive
evolved stars, Luminous Blue Variables (LBVs). These stars are thought
to be the link between the main sequence O stars and the final
products of massive-star evolution such as Wolf-Rayet stars and
Supernovae (see Lamers et al 2001).  Their large scale structures have
bipolar morphologies (Nota et al 1995). However, it remains unclear
whether these bipolar nebulae are due to spherically symmetric winds
interacting with a pre-existing density contrast or whether the star
is undergoing enhanced mass loss in the equatorial plane, perhaps due
to rotation (Dwarkadas \& Balick 1998; Dwarkadas \& Owocki 2002).  The
presence of an equatorial density contrast, or a disk, can potentially
answer questions related to the nature of the formation of the bipolar
flows.  To detect these disks, Davies, Oudmaijer \& Vink (2005)
undertook spectropolarimetric observations of a large sample of both
Galactic and Magellanic Cloud Luminous Blue Variables.  Around half of
the objects show a line effect.  As an example AG Car is shown in
Fig.~\ref{lbv}. Schulte-Ladbeck et al. (1994) already observed this
line effect and also discovered polarization variability in AG Car. As
the points in {\it QU} space followed a straight line in the data at
their disposal, the presence of a disk was readily inferred. New data
are added to this dataset and presented in Fig.~\ref{lbv} as well. The
big cross in the middle is the polarization at the line centre. It is
not variable and most likely represents the interstellar
polarization. This means that the intrinsic polarization angle varies
randomly with time, opposed to what would be expected from a
circumstellar disk and illustrating that the addition of more data
here complicated the issue.

Davies et al. (2005) discuss several mechanisms that may explain the
variability, and arrive at the random ejection of clumps in the wind
as the most plausible scenario. In fact, this was introduced earlier
in the case of P Cygni by Nordsieck et al (2001 - using results by
Taylor et al. 1991), where the variability, at smaller magnitude, was
also found to be random. There are not enough variability data
available to extrapolate this particular conclusion to the entire
class.  However, the main result that roughly half of the LBVs show a
line effect and thus the evidence for either disks or clumpiness is
robust. For comparison, Harries et al (1998) find that only 10\% of
Wolf-Rayet stars and 25\% of O supergiants show the effect (Harries et
al. 2002), while the majority of evolved B[e] supergiants do
(Schulte-Ladbeck et al 1993; Oudmaijer \& Drew 1999). Pending further
analysis, the data may hint at different evolutionary sequences for WR
stars and LBVs and further study is warranted. As clumps seem to
dominate at least the spectropolarimetry of massive evolved stars, it
would turn out that John Dyson had indeed been right to warn the
unsuspecting community of more complexities!

\section{Outlook}

\begin{figure}[t]
\centering
 \mbox{\epsfxsize=0.6\textwidth\epsfbox[1 600 300 800]{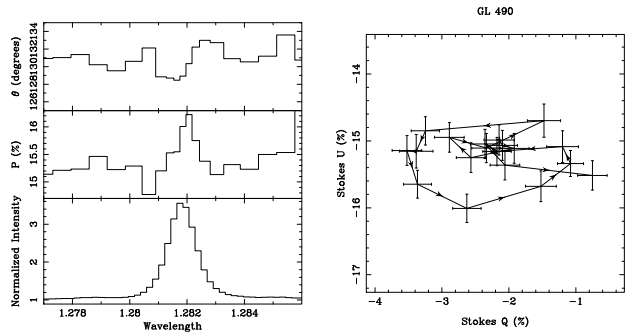}}
\caption{
Pa$\beta$ polarization data of AFGL 490.  The left hand graph shows
the polarization data, as in Figure~1, now as a function of 
wavelength in
$\mu$m. The bottom panel shows the (normalized) intensity spectrum,
the data are rebinned to a corresponding accuracy in the
polarization of 0.25\% The right hand plot shows the Stokes QU vectors
with the same binning applied. Data from Oudmaijer et al. (2005).
\label{mww} }
\end{figure}


The previous sections discussed H$\alpha$ data and the application of
the method to optically visible objects.  The earliest type Herbig Be
stars sample the highest mass stars already, but even with the advent
of the 8m class telescopes, the deeply embedded massive Young Stellar
Objects remain elusive.  To be able to determine whether (accretion)
disks around such objects are present we need study them at longer
wavelengths where the objects are brighter. Their infrared hydrogen
recombination lines are fairly strong (Bunn et al 1995) and would be
excellent target lines for spectropolarimetry. Little medium
resolution spectropolarimetry has been performed in the near-infrared,
so far however.  Here we present some of the very first data across
atomic lines of a stellar object.  Oudmaijer et al. (2005) undertook
proof-of-concept observations at UKIRT and observed amongst others the
massive YSO AFGL 490. The results around the Pa$\beta$ line are shown
in Figure~\ref{mww}. The data show a clear line-effect, and are
consistent with Schreyer et al. (2006) who, in parallel, observed a
larger scale rotating disk around the object.

Going to the other end of the evolution of massive stars, much
progress has been made over the last years in the study of the
geometry of supernova ejecta using spectropolarimetry. Studies from
e.g.  Wang et al. (2003) and Leonard et al. (2005 - see references
herein), building on the original suggestion by McCall (1984) find
large intrinsic polarizations, and thus asymmetries, towards
core-collapse supernovae. Ultimately, data such as these, and combined
with those on their progenitors such as the LBVs mentioned earlier,
will provide strong observational constraints on not only the
supernova mechanism itself, but also on the possible Gamma ray
bursters resulting from them.

In this review, we have discussed the enormous progress that has been
made, both observationally and theoretically, over the past decade in
the field of linear spectropolarimetry and its application to star
formation and stellar evolution.  With the coming of age of
spectrpolarimetry on 8-m class telescopes and improved modelling, the
future hold even more exciting prospects in store.

\subsection*{Acknowledgements} 
It is a pleasure to thank John Dyson for his inspiration, not only as
an astronomer, but also as a friend, and I am happy to have been able
to work in Leeds with him. My sincere thanks to my long-term
collaborators in spectropolarimetry, Janet Drew, Jorick Vink and Tim
Harries. Joe Mottram and Ben Davies are thanked for their comments on
an earlier draft of the manuscript. Jon Bjorkman is thanked for
providing figure~\ref{bjorkmod}.

\section{References}

Adams F.C. 1993, ASP Conf. Ser. 35, p.56  \\
Akeson R.L., Walker C.H., Wood, K. et al. 2005, ApJ 622, 440 \\ 
Bally J., Zinnecker H. 2005,  AJ 129, 2281\\ 
Baines D. Oudmaijer R.D., Porter J.M., Pozzo M., 2006  MNRAS 367, 737\\
Bertout C. 1989 ARA\&A 27, 351 \\
 Bonnell I.A., Bate M.R.,  Zinnecker H.  1998, MNRAS 298, 93\\
 Bunn J., Hoare M.G., Drew J.E. 1995, MNRAS 272,  346 \\
 Carciofi A.C., Bjorkman J.E. 2006, ApJ 639, 1081 \\
 Cassinelli J.P., Nordsieck K.H., Murison  M.A. 1987, ApJ 317, 290\\
 Clarke C.J. 2001, IAUS 200, 346\\
 Davies B., Oudmaijer R.D., Vink J.S. 2005, A\&A 439, 1107 \\
 Dougherty  S.M. \& Taylor R. 1992, Nature 359, 808 \\
 Dullemond C.P., van den  Ancker M.E., Acke B., van Boekel R. 2003, ApJ 594, L47 \\
 Dutrey A., Guilloteau S., Prato L., Simon M., Duvert G., Schuster K., Menard F. 1998 A\&A 338, L63 \\
Dwarkadas  V.V. and Balick B. 1998, AJ 116, 829 \\
 Dwarkadas V.V. and Owocki  S. 2002, ApJ 581, 1337 \\
 Dyson J.E. 1991 in ``Mass loss on the AGB  and beyond'', ed. H.E. Schwarz, ESO Conference and Workshop  Proceedings 46, p. 1 \\
 Fukagawa M., Tamura M., Itoh Y., Hayashi  S.S., Oasa Y.2003, ApJ, 590, 49 \\
 Grady C.A., Polomski E.F., Henning  Th. et al., 2001, AJ, 122, 3396 \\
 Grinin V.P., Th\'e P.S., de Winter  D., Giampapa A.N., Tambovtseva L.V., van den Ancker M.E. 1994, A\&A  292, 165 \\
 Harries T.J., Hillier D.J., Howarth I.D. MNRAS 296, 1072 \\
 Harries  T.J. 2000, MNRAS 315, 722 \\
 Harries T.J., Howarth I.D., Evans C.J.  2002, MNRAS 337, 341 \\
Hoare M.G., Drew J.E., Muxlow T.B., Davis R.J. 1994 ApJ 421, L51 \\
Hoare M.G. 2002 ASP Conf Proc, Vol. 267, p 137 \\
 Hubrig S., Sch\"oller M., Yudin R.V. 2004,  A\&A 428, 1\\
 Hubrig S., Sch\"oller M., Yudin R.V., Pogodin  M.A. 2006, A\&A 446, 1089\\
 Jiang Z., Tamura M., Fukagawa M., Hough J., Lucas P.,  Suto H., Ishii M., Yang J. 2005 Nature 437 112 \\
Johns-Krull C.M., Valenti J.A., Koresko C. 1999, ApJ 516, 900 \\
 Kaye A.B., Gies  D.R. 1997, ApJ 482, 1028 \\
 Lamers H.J.G.L.M., Nota A., Panagia N., Smith L.J., Langer N.  2001, ApJ 551, 764 \\
Leinert C., Richichi A., Haas M. 1997, A\&A 318, 472  \\
Leonard D.C., Weidong L., Filippenko A.V., Foley R.J., Chornock R. 2005, ApJ 632, 450 \\
 Mannings V. \& Sargent A. I., 2000, ApJ, 529, 391\\
 McCall M.L. 1984, MNRAS 210, 829 \\
 McDavid D. 1999,  PASP 111, 494 \\
Mora A., Eiroa C., Natta A. et al. 2004 A\&A 419, 225\\
Mottram J.C., Vink J.S., Oudmaijer R.D., Patel M. 2007, MNRAS submitted \\
Muzerolle J. Calvet N., Hartmann L. 2001, ApJ 550, 944 \\
 Norberg P.,  Maeder A. 2000, A\&A 359, 1035\\
 Nordsieck K.H. et al. 2001 in ``P  Cygni 2000: 400 years of progress'' ed. M. de Groot \& C. Sterken, ASP Conf. Ser. 233, 261 \\
 Nota A.,Livio M., Clampin M., Schulte-Ladbeck R. 1995, ApJ 448, 788 \\
 Oudmaijer R.D. and Drew J.E. 1999, MNRAS 305, 166\\
 Oudmaijer R.D.,  Proga D., Drew J.E., de Winter D. 1998, MNRAS, 300, 170\\
 Oudmaijer R.D., Palacios J., Eiroa C. et al. 2001, A\&A 379, 564 \\
 Oudmaijer  R.D. Drew J.E., Vink J.S. 2005, MNRAS 364, 725\\
 Patel N.A.,  Curiel S., Sridharan T.K., Zhang Q., Hunter T.R., Ho P.T.P.,  Torrelles J.M., Moran J.M., G\'omez J.F., Anglada G. 2005 Nature 437, 109 \\
 Poeckert R. and Marlborough J.M. 1976, ApJ 206, 182 \\
 Porter J.M., Rivinius  T. 2003, PASP 115, 1153 \\
 Quirrenbach A., Buscher D.F., Mozurkewich  D., Hummel C.A., Armstrong J.T. 1994, A\&A 283, L13\\ 
 Quirrenbach  A., Bjorkman K.S., Bjorkman J.E. et al 1997, ApJ 479, 477\\
 Schreyer K., Semenov D., Henning Th., Forbrich J. 2006, ApJ 637, L129 \\
 Schulte-Ladbeck R.E., Shepherd D.S., Nordsieck K.H., et al. 1992, ApJ  401, L195 \\
 Schulte-Ladbeck R.E.,Leitherer C., Clayton G.C.  et al. 1993, ApJ 407, 723 \\
 Schulte-Ladbeck R.E., Clayton G.C., Hillier D.J., Harries T.J., Howarth I.D. 1994, ApJ 429, 846 \\
 Serkowski K., 1962,  in Adv. Astron. Astroph. 1, 289 \\
 Shepherd D.S., Claussen M.J., Kurtz S.E. 2001, Science 292, 1513 \\
Taylor M., Nordsieck K.H.,  Schulte-Ladbeck R.E., Bjorkman K.S. 1991, AJ 102, 1197 \\
 Th\'{e} P.S., de Winter D., Perez M.R., 1994, A\&AS, 104, 315\\
 Vink J.S., Drew J.E., Harries T.J.,  Oudmaijer R.D. 2002, MNRAS, 337, 356\\ 
 Vink J.S., Drew J.E., Harries  T.J., Oudmaijer R.D., Unruh Y. 2003, A\&A 406, 703 \\
 Vink J.S., Drew  J.E., Harries T.J., Oudmaijer R.D., Unruh Y. 2005a, MNRAS, 359, 1049\\
 Vink J.S., Harries T.J., Drew J.E. 2005b, A\&A 430, 215 \\
 Wade G.A.,  Drouin D., Bagnulo S., Landstreet J.D., Mason E., Silvester J.,  Alecian E., B\"ohm T., Bouret J.-C., Catala C., Donati J.-F. 2005,  A\&A 442, L31 \\
Wang L., Baade D., H\"offlich P., Wheeler J.C. 2003, ApJ 592, 457 \\
 Waters L.B.F.M., Waelkens C., Trans N.R.  1991 in  ``Mass loss on the AGB and beyond'', ed. H.E. Schwarz, ESO Conference  and Workshop Proceedings 46, p. 298 \\
 Wolf S., Stecklum B., Henning  Th. 2001, IAUS 200, 295\\
 Wolfire M.G., \& Cassinelli J.P. 1987, ApJ  319, 850\\
Wood K., Brown J.C., Fox G.K. 1993, A\&A 271, 492\\
 Wood K., Bjorkman K.S., Bjorkman J.E. 1997, ApJ 477, 926  \\
 Yorke H.W. and Kruegel E. 1977, A\&A 54, 183\\
 Yorke H.W. and Sonnhalter C. 2002, ApJ 569, 846\\


\end{document}